\documentclass[fleqn,usenatbib]{mnras}

\usepackage{newtxtext,newtxmath}

\usepackage[T1]{fontenc}

\DeclareRobustCommand{\VAN}[3]{#2}
\let\VANthebibliography\thebibliography
\def\thebibliography{\DeclareRobustCommand{\VAN}[3]{##3}\VANthebibliography}

\usepackage{graphicx}	
\usepackage{amsmath}	

\newcommand{\NIIl}{{[N\,{\sc ii}]\,$\lambda$}}
\newcommand{\NIII}{{N\,{\sc iii}]}}
\newcommand{\NIIIl}{{N\,{\sc iii}]\,$\lambda$}}
\newcommand{\NIVl}{{N\,{\sc iv}]\,$\lambda$}}
\newcommand{\NIV}{{N\,{\sc iv}]}}
\newcommand{\CIV}{{C\,{\sc iv}]}}

\newcommand{\OIIIl}{{[O\,{\sc iii}]\,$\lambda$}}

\newcommand{\NeIIIl}{{[Ne\,{\sc iii}]\,$\lambda$}}

\newcommand{\CIII}{{C\,{\sc iii}]}}

\newcommand{\CIIIll}{{C\,{\sc iii}] $\lambda\lambda$}}

\newcommand{\OIIIsfll}{{O\,{\sc iii}]\,$\lambda$$\lambda$}}
\newcommand{\OII}{{[O\,{\sc ii}]}}

\newcommand{\OIIll}{{[O\,{\sc ii}]\,$\lambda\lambda$}}

\newcommand{\HeII}{{He\,{\sc ii}\,}}

\defcitealias{Bunker2023}{B23}

\title[Nitrogen enhancement in GN-z11]{
Nitrogen enhancements 440~Myr after the Big Bang: super-solar N/O, a tidal disruption event or a dense stellar cluster in GN-z11?
}

\author[A. J. Cameron et al.]{
Alex J. Cameron,$^{1}$\thanks{E-mail: alex.cameron@physics.ox.ac.uk} Harley Katz$^{1}$, Martin P. Rey$^{1}$ and Aayush Saxena$^{1,2}$
\\
$^{1}$Department of Physics, University of Oxford, Denys Wilkinson Building, Keble Road, Oxford, OX1 3RH, UK \\
$^{2}$Department of Physics and Astronomy, University College London, Gower Street, London WC1E 6BT, UK \\
}

\date{Accepted XXX. Received YYY; in original form ZZZ}

\pubyear{2023}

\begin{document}
\label{firstpage}
\pagerange{\pageref{firstpage}--\pageref{lastpage}}
\maketitle

\begin{abstract}
Recent observations of GN-z11 with \emph{JWST}/NIRSpec revealed numerous oxygen, carbon, nitrogen, and helium emission lines at $z=10.6$. Using the measured line fluxes, we derive abundance ratios of individual elements within the interstellar medium (ISM) of this super-luminous galaxy. Driven by the unusually-bright \NIIIl1750 and \NIVl1486 emission lines (and by comparison faint \OIIIsfll1660, 1666 lines), our fiducial model prefers $\log({\rm N/O})>-0.25$, greater than four times solar and in stark contrast to lower-redshift star-forming galaxies.
The derived $\log({\rm C/O})>-0.78$, ($\approx$30 \% solar) is also elevated with respect to galaxies of similar metallicity ($12+\log({\rm O/H})\approx7.82$), although less at odds with lower-redshift measurements. 
Given the long timescale typically expected to enrich nitrogen with stellar winds, traditional scenarios require a very fine-tuned formation history to reproduce such an elevated N/O. We find no compelling evidence that nitrogen enhancement in GN-z11 can be explained by enrichment from metal-free Population~III stars. Interestingly, yields from runaway stellar collisions in a dense stellar cluster or a tidal disruption event provide promising solutions to give rise to these unusual emission lines at $z=10.6$, and explain the resemblance between GN-z11 and a nitrogen-loud quasar. These recent observations showcase the new frontier opened by \emph{JWST} to constrain galactic enrichment and stellar evolution within 440~Myr of the Big Bang.

\end{abstract}




\section{Introduction} \label{sec:intro}

Chemical abundance measurements provide powerful constraints on the physical mechanisms underlying galaxy formation and evolution.
Elements heavier than hydrogen and helium (metals) are formed via processes associated with the stellar life cycle, and the assembly of each galaxy is hence inherently linked with chemical enrichment (see \citealt{Maiolino2019} for a review). One way this connection manifests is through the mass-metallicity relation. The correlation between metal enrichment (metallicity) and galaxy stellar mass has been well established across the history of the Universe, both for metals in the gas-phase (\citealt{Lequeux1979, Tremonti2004, Mannucci2010, Andrews2013, Steidel2016, Yates2020, Sanders2021}) and for metals locked in stars (\citealt{Gallazzi2005, Kirby2013, Zahid2017, Cullen2019, Kashino2022}). These studies demonstrate a general trend whereby metal enrichment tracks star formation across the Universe, with more evolved galaxies being more enriched with metals, and higher-redshift galaxies having lower metallicities (e.g. \citealt{Maiolino2019} for a summary).

Studies of individual heavy elements and their relative abundance ratios with respect to each other can provide further constraints on galaxy evolution. Certain elements are formed by distinct astrophysical channels that occur on different timescales (see e.g. \citealt{Kobayashi2023} for a review). The relative abundances of metals formed through these different channels will thus vary as a galaxy evolves. Hence, the relative abundances of heavy elements can reveal the underlying physical process that dominated the growth of a galaxy.

Of particular interest are oxygen, carbon and nitrogen. These three elements are amongst the most abundant metals in the Universe. They can be readily observed in the ISM of galaxies via prominent emission lines, and they are formed preferentially by specific astrophysical processes with distinct enrichment timescales. Oxygen is predominantly formed in core-collapse supernovae (CCSN) that occur on short timescales following the onset of star-formation (see e.g. \citealt{Nomoto2013} for a review). Moderate levels of carbon and nitrogen are formed in CCSN, but these two metals are also enriched via stellar winds during the asymptotic giant branch (AGB) phase of intermediate mass stars (see e.g. \citealt{Karakas2014} for a review). Since such intermediate-mass stars have longer main-sequence lifetimes before their giant phase, enrichment in carbon and nitrogen is expected to occur on longer timescales, with nitrogen potentially lagging behind carbon. Thus, the canonical picture of chemical evolution (see e.g. discussion in \citealt{Kobayashi2020}) is that galaxies are rapidly enriched with oxygen (and other $\alpha$-elements) following a burst of star formation, while the nitrogen and carbon content of a galaxy slowly grows as the stellar populations age. Emission-line galaxies represent an ideal laboratory to quantitatively test these galactic chemical evolution models.

Collisionally-excited emission lines (CELs), particularly \OIIIl5007 and \OIIll 3726, 3729 arising from ionised oxygen, have been extensively used to derive gas-phase oxygen abundances (O/H) in the ISM of galaxies (e.g. \citealt{Andrews2013, Curti2020, Sanders2021}) and now extend to $z\gtrsim7$ thanks to the \emph{JWST} \citep{Curti2023, Tang2023, Nakajima2023}. The gas-phase nitrogen abundance, and its ratio to oxygen (N/O) is typically probed with the [N~{\sc ii}] $\lambda$6583 emission line. Such studies have revealed that N/O is well-correlated with O/H both in the local Universe and at higher redshift \citep{Pilyugin2012, PerezMontero2013, Masters2014, Amorin2017, Berg2020, HaydenPawson2022}, with a characteristic shape showing a plateau below $12+\log({\rm O/H})\approx8.1$ and a steady increase toward higher metallicities. High N/O ratios are typically only found in galaxies with super-solar metallicities, consistent with the expectation of slow nitrogen enrichment over the course of the Universe, while low-metallicity galaxies tend to stay well below $\log ({\rm N/O}) \lesssim -0.5$.
 
A similar method can be used to derive gas-phase carbon abundances using rest-frame ultraviolet (UV) emission to estimate the evolution of C/O over time (e.g. \citealt{Garnett1995, Garnett1999, Berg2016, Berg2019, Steidel2016, Llerena2022} and references therein). Such studies find a comparable picture to that of nitrogen: galaxies with higher O/H also exhibit higher C/O. Measurements of C/O have been extended to $z\geq 6$ galaxies with JWST, providing evidence that carbon enrichment has already proceeded beyond that expected from CCSN alone \citep{ArellanoCordova2022, Jones2023}.

Recently, \citet{Bunker2023} (\citetalias{Bunker2023} hereafter) reported a \emph{JWST}/NIRSpec spectrum for GN-z11. This galaxy was first identified as a high-redshift candidate in \citep{Oesch2015}, was tentatively confirmed to have a redshift of $z=11.1$ based on \emph{HST} grism spectroscopy of the Lyman-$\alpha$ break (\citealt{Oesch2016}), and is now confirmed unambiguously at $z=10.60$ (\citealt{Bunker2023}).
GN-z11 is remarkably luminous compared to the $z\sim10-11$ luminosity function (see \citealt{Robertson2022} for a review and e.g. \citealt{Bouwens2022, Finkelstein2022, Harikane2022, PerezGonzalez2023} for recent determinations with the JWST). Of further intrigue, the NIRSpec spectrum published in \citetalias{Bunker2023} also shows numerous emission lines, arising from both oxygen (\OIIIl4363, \OIIll3726, 3729 and tentative \OIIIsfll1660, 1666) and carbon ([C {\sc iii}] $\lambda$1907 + C {\sc iii}] $\lambda$1909\footnote{For brevity, we will hereafter refer to [C {\sc iii}] $\lambda$1907 + C {\sc iii}] $\lambda$1909 as C {\sc iii}] $\lambda\lambda$1909.}, and tentative \CIV). Even more surprisingly, unusually-bright nitrogen emission is detected (\NIIIl1750, \NIVl1486) with the measured line fluxes being higher than that measured for oxygen in the rest-frame UV.

\NIIIl 1750\footnote{We note that this \NIII\ emission feature consists of a quintet of emission lines between rest-frame 1746~\AA{} and 1755~\AA{}. Throughout this paper we refer to the sum of this complex as \NIIIl 1750.} is rarely observed in galaxy spectra, and when detected it is typically much fainter than other nearby rest-frame UV lines (particularly \OIIIsfll1660,1666). This is verified both in local analogues to high-redshift galaxies (e.g. \citealt{Mingozzi2022}), individual galaxies at $z\sim2$ \citep{Berg2018}, stacked spectra of $z\sim2-4$ galaxies \citep{LeFevre2019, Saxena2022}, and in a tentative detection at $z>7$ with \emph{JWST}/NIRISS \citep{RobertsBorsani2022}. A small subset of SDSS quasars exhibit prominent \NIIIl 1750 emission (so called `nitrogen-loud' quasars; \citealt{Jiang2008, Batra2014}), but \citetalias{Bunker2023} do not find any unambiguous signature of AGN activity in GN-z11 (which we revisit in the discussion of Section~\ref{sec:discussion:AGN}).

A possible interpretation to explain such bright nitrogen emission lines is an unusually high nitrogen content, and an elevated N/O ratio. Indeed, this is suggested by \citetalias{Bunker2023}. In this paper, we use the emission line flux ratios published by \citetalias{Bunker2023} to quantify the N/O and C/O abundance ratios in GN-z11 (Section~\ref{sec:abundances}). Our fiducial model implies super-solar N/O and elevated C/O at $z=10.6$ where the age of the Universe is less than 500 Myr. We discuss in Section~\ref{sec:discussion} that these abundance ratios are challenging to explain with traditional enrichment arguments where nitrogen and carbon are enriched by stellar evolution channels on long timescales, and review other more exotic scenarios that could explain such elevated values at $z=10.6$. We present a summary in Section~\ref{sec:conclusion}. 


\section{Abundance Measurements} \label{sec:abundances}

In this section we outline our methods for deriving limits on a series of ion abundance ratios in GN-z11. We use these calculations to place constraints on the N/O, C/O and O/H abundance ratios in GN-z11, which we summarise in Tables~\ref{tab:fiducial} and \ref{tab:abund}. Discussion of the implications of these derived values can be found in Section~\ref{sec:discussion}. 

Emissivity calculations in this section are performed with {\sc pyneb} \citep{Luridiana2015_pyneb}, using the atomic data from the {\sc chianti} database (version 10.0.2; \citealt{Dere1997_chianti, DelZanna2021_chianti}).
Emission lines fluxes used in these calculations are taken from Table~1 in \citetalias{Bunker2023}, adopting measurements from their medium-resolution grating spectra. However, the \OIIIl4363 line is only detected in the prism spectrum, for which we note all reported fluxes are systematically lower. Thus, we scale the reported \OIIIl4363 prism flux to match those of the grating using the nearby H$\gamma$ line, which is reported in both the prism and the grating.
Abundance ratios are calculated from reported flux ratios and estimated ratios of emission line emissivities. 
For example, for N$^{++}$/O$^{++}$, we have
\begin{equation}
    \frac{\text{N}^{++}}{\text{O}^{++}} = \frac{f_{\rm N\textsc{iii}] 1750}}{f_{\rm O\textsc{iii}] 1660, 1666}} \times \frac{\epsilon_{\rm O\textsc{iii}] 1660, 1666}}{\epsilon_{\rm N\textsc{iii}] 1750}},
\label{eq:example_ratio}
\end{equation}
where $\epsilon_{x}$ is the emissivity of each emission line that depends on the electron temperature, $T_e$, and density, $n_e$.

\subsection{Electron temperature constraints}
\label{sub:temperature}

\begin{table}
    \centering
    \begin{tabular}{ccc}
\hline
Abundance ratio & Fiducial & Conservative \\
\hline
$\log({\rm N/O})$ & $>-0.25$ & $>-0.49$ \\
$\log({\rm C/O})$ & $>-0.78$ & $>-0.95$ \\
$12+\log({\rm O/H})$ & $7.82$ & $<8.6$ \\
\hline
    \end{tabular}
    \caption{Summary of the abundance limits derived in Section~\ref{sec:abundances}. The `fiducial' column takes its values from the $T_e=1.46\times10^4\ {\rm K}$ column of Table~\ref{tab:abund}. The N/O and C/O ratios in the `conservative' column are the lowest values obtained from any combination of assumptions in Table~\ref{tab:abund}, excluding the strongly disfavoured $T_e=3\times10^4\ {\rm K}$. The `conservative' O/H value adopts the highest O/H value from Table~\ref{tab:abund} as an upper limit. Note that O/H abundance ratios are much more sensitive to modelling assumptions than metal abundance ratios. For reference, solar values are: $\log ({\rm N/O})_\odot = -0.86$, $\log ({\rm C/O})_\odot = -0.26$, $12+\log ({\rm O/H})_\odot = 8.69$ \citep{Asplund2009}.}
    \label{tab:fiducial}
\end{table}

Because \OIIIl5007 is beyond the wavelength coverage of NIRSpec at $z=10.6$, we first use \OIIIsfll 1660, 1666 along with \OIIIl4363 to derive an electron temperature constraint. If we assume that there is no dust reddening in the system (consistent with the H$\delta$/H$\gamma$ ratio reported), the $2\sigma$ upper limit on the \OIIIsfll 1660,1666/\OIIIl~4363 ratio ($<1.57$) gives us an upper limit on the temperature of $<1.25\times 10^4$~K. This value is lower than previous electron temperature measurements at this epoch \citep[e.g.][]{Curti2023, Katz2023_ero}, and also that expected from the reported presence of \NIV\ emission. This low electron temperature is however consistent with the galaxy being at higher metallicity compared to other high-redshift objects.

We consider an alternative approach for estimating the electron temperature, making use of the \OIIIl4363/\NeIIIl3869 ratio.
Neon and oxygen are both $\alpha$-elements and the Ne/O abundance ratio has been observed to be quite consistent across a large range of abundances and redshifts \citep[e.g.][]{Berg2019_Ne, Berg2020, ArellanoCordova2022}.
Furthermore, Ne$^{++}$ traces a similar ionisation zone to O$^{++}$, meaning that \NeIIIl3869 / \OIIIl5007 flux ratios typically do not show large variations \citep[e.g.][]{Witstok2021}.
Thus, we solve for the temperature at which the measured \OIIIl4363 / \NeIIIl3869 flux ratio reproduces the solar Ne/O abundance ratio. This results in a temperature of $T_e=1.46\pm0.26 \times 10^4$ K, which is in reasonable agreement with the temperature limit inferred from the \OIIIsfll 1660,1666/\OIIIl 4363 ratio.
We adopt this temperature for our fiducial calculation.
Assuming half-solar and twice-solar Ne/O abundance ratios yield $T_e=1.05\times10^4$ K and $2.36\times10^4$ K respectively.
To bracket the range of temperatures implied by this spectrum, we perform our abundance analysis also at these temperatures and finally at $3.0\times10^4$ K as a final bounding case\footnote{Temperatures of $\sim3\times 10^4$ K have been reported in $z>8$ galaxies \citep[e.g.][]{Katz2023_ero}}, although this scenario is not favoured.

Throughout this calculation we assume these quoted temperatures apply to the `high-ionisation' zone, (i.e., $T_e$(O{\sc iii}) = $T_e$(N{\sc iii}) = $T_e$(C{\sc iii})). We assume that O$^{+}$ traces a different ionisation zone with different $T_e$.
We consider two conversions from $T_e$(O{\sc iii}) to $T_e$(O{\sc ii}): (i) the calibration provided in \citet{Pilyugin2009} and a more exaggerated case where $T_e$(O{\sc ii}) $=0.7\times T_e$(O{\sc iii}).
We initially adopt $n_e=100$ cm$^{-3}$, although the effect of density variations is also discussed below.
Abundance ratios derived for these different temperatures are provided in full in Table~\ref{tab:abund}.

\begin{figure*}
    \centering
    \includegraphics[width=\textwidth]{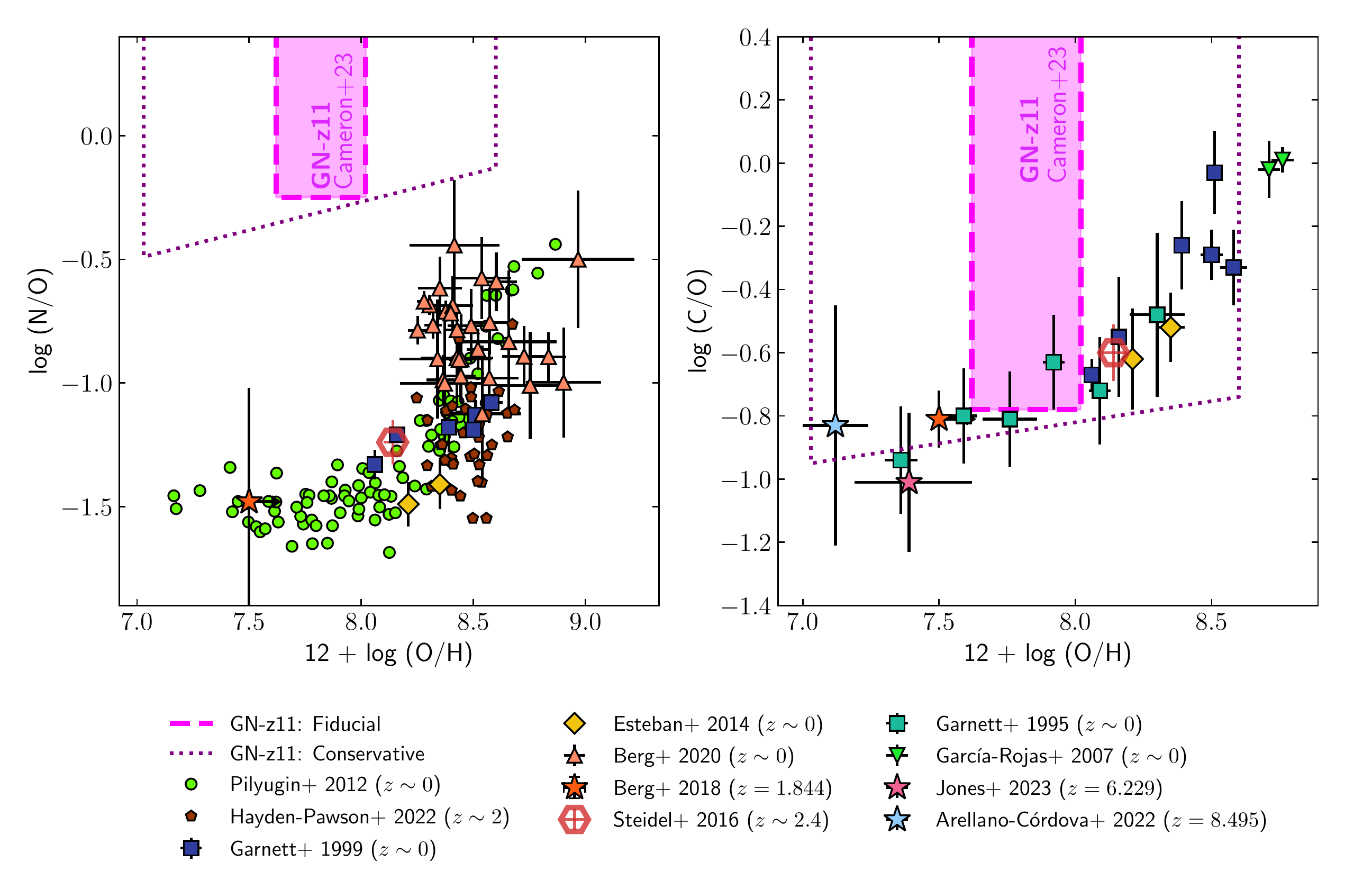}
    \caption{
    Pink shaded regions show the range of abundance ratios for GN-z11 implied by our fiducial model (dashed) and our more conservative assumptions (dotted).
    \textit{Left}: Nitrogen-to-oxygen abundance ratio compared to total oxygen abundance. We show comparison samples of $z\sim0$ H{\sc ii} regions (green circles from \citealt{Pilyugin2012}; blue squares from \citealt{Garnett1999}; yellow diamonds from \citealt{Esteban2014}; orange triangles from \citealt{Berg2020}), $z\sim2$ galaxies from \citet{Berg2018} (orange star) and \citet{HaydenPawson2022} (brown pentagons), and the $z=2.4$ composite spectrum from \citet{Steidel2016} (red hexagon). Our inferred N/O for 
    GN-z11 is highly super-solar and unlike lower-redshift galaxies.
    \textit{Right:} Carbon-to-oxygen abundance ratio compared to total oxygen abundance. We show comparison samples of $z\sim0$ H{\sc ii} regions (turquoise squares from \citealt{Garnett1995}; blue squares from \citealt{Garnett1999}; green triangles from \citealt{GarciaRojas2007}; yellow diamonds from \citealt{Esteban2014}).
    The $z=2.4$ composite from \citet{Steidel2016} is shown by the red hexagon.
    We show two measurements of C/O in individual galaxies with \emph{JWST}/NIRSpec: a galaxy at $z=6.229$ from \citet{Jones2023} (pink star) and a galaxy at $z=8.495$ from \citet{ArellanoCordova2022} (light blue star).
    }
    \label{fig:fig1}
\end{figure*}

\subsection{Constraints on N/O in GN-z11}
\label{sub:measure_no}

We now turn to exploring the possible N/O ratios implied by the emission line measurements reported by \citetalias{Bunker2023} for GN-z11.
With measurements reported for the \NIIIl 1750 and \NIVl 1486 emission lines, we can sample the N$^{++}$ and N$^{3+}$ ions. Fluxes (or limits) are reported for three oxygen lines: \OIIIsfll 1660, 1666, \OIIll3726, 3729, and \OIIIl4363. The \NIIl 6583 emission line is not within the spectral coverage which has historically been the most common way of determining nitrogen abundances \citep[e.g.][]{PerezMontero2017}.
Although it is not a widely used tracer, nitrogen abundance constraints have been reported from \NIIIl 1750 at $z\sim0$ \citep{Garnett1999} and $z\sim2$ \citep{Berg2018}.

We first consider N$^{++}$/O$^{++}$ from the \NIIIl 1750 / \OIIIsfll 1660, 1660 ratio. Since the \OIIIsfll 1660, 1660 emission is reported as an upper limit, for a given adopted temperature, this ratio provides a lower limit on the N$^{++}$/O$^{++}$ ratio. With our fiducial temperature we derive a value of $\log({\rm N^{++}/O^{++}})>-0.07$. Considering our range of temperatures (Section~\ref{sub:temperature}), we obtain ion abundance ratios between $-0.12\leq$ log(N$^{++}$/O$^{++}$) $\leq 0.1$, exhibiting only a mild $T_e$ dependence.

We can also derive an estimate of N$^{++}$/O$^{++}$ from the \NIIIl1750~/~\OIIIl4363 ratio. This has the disadvantage of a much larger wavelength difference, making the ratio sensitive to any wavelength-dependent dust corrections. Assuming there is no dust, we find log(N$^{++}$/O$^{++})=-0.19$ for our fiducial temperature. We note that invoking the presence of dust preferentially increases the \NIIIl1750 / \OIIIl4363 flux ratio, which serves to increase the inferred N$^{++}$/O$^{++}$. Considering our range of temperature, we find that, derived from \NIIIl 1750 / \OIIIl4363, the N$^{++}$/O$^{++}$ ratio has a somewhat larger $T_e$-dependence, and that the dependence is opposite, varying between $-0.48\leq$ log(N$^{++}$/O$^{++}$) $\leq 0.04$ (Table~\ref{tab:abund}).

The emissivity ratio of $\epsilon_{\rm O\textsc{iii}] 1660, 1666}/\epsilon_{\rm N\textsc{iii}] 1750}$ is essentially constant with density up to $\sim10^5$ cm$^{-3}$, suggesting that the impact of density variations is minor. A small density dependence appears at $n_e \gtrsim 10^6$ cm$^{-3}$, although not significant enough to appreciably reduce the derived N$^{++}$/O$^{++}$. Furthermore, the $n_e$-dependence of $\epsilon_{\rm [O\textsc{iii}] 4363}/\epsilon_{\rm N\textsc{iii}] 1750}$ implies {\em higher} N$^{++}$/O$^{++}$ at high density, further disfavouring this solution.

Although N$^{++}$ and O$^{++}$ are both high-ionisation ions and will trace similar regions of the nebula, we cannot necessarily assume N/O = N$^{++}$/O$^{++}$.
Since the \NIIIl1750 line has not been widely studied in the literature, ionisation correction factors (ICF) for the N$^{++}$/O$^{++}$ ratio are typically not considered \citep[e.g.][]{PerezMontero2017, Amayo2021}. The second and third ionisation energies of nitrogen (29.6 eV and 47.4 eV) and oxygen (35.1 eV and 54.9 eV) imply that the nebular zone probed by emission from N$^{++}$ ions should contain both O$^{+}$ and O$^{++}$ ions \citep{NIST_ASD}. We therefore assume that N$^{++}$/(O$^{+}$ + O$^{++}$) provides a lower limit on the total N/O abundance ratio, and thus we derive the N$^{++}$/O$^{+}$ ratio from the detected \OIIll3726, 3729 lines.

Unlike the N$^{++}$/O$^{++}$ calculation, we do not assume that $T_e$(O{\sc ii}) = $T_e$(N{\sc iii}).
We do not have any direct constraints on the temperature of the low-ionisation zone, so we instead derive $T_e$(O{\sc ii}) for each assumed temperature using the relation from \citet{Pilyugin2009}, yielding $T_e$(O{\sc ii}) = [1.14, 1.48, 2.24, 2.77] $\times 10^4$ K.

Assuming this two-zone model, we find $0.38\leq$ log(N$^{++}$/O$^{+}$) $\leq 1.69$, consistent with the finding from \citetalias{Bunker2023} that GN-z11 has a very highly ionised ISM.
We note that such $T_e$(O{\sc ii})-$T_e$(O{\sc iii}) calibrations are highly uncertain \citep{Yates2020, Cameron2021}. Given that lower $T_e$(O{\sc ii}) leads to lower \OII\  emissivity, and thus lower N$^{++}$/O$^{+}$, we repeat this calculation with an exaggerated case where $T_e$(O{\sc ii}) = $0.7\times T_e$(O{\sc iii}). This yields slightly lower values between $0.23\leq$ log(N$^{++}$/O$^{+}$) $\leq 0.96$, which we include as part of our conservative model. 

As for N$^{++}$ and O$^{++}$, dust corrections might need to be applied for GN-z11, but we note that they would only increase the derived N$^{++}$/O$^{+}$, and thus N/O. Furthermore, high densities are once again disfavoured. The derived N$^{++}$/O$^{+}$ shows limited density dependence up to $\lesssim10^5$ cm$^{-3}$, above which the emissivity of \OIIll 3726, 3729 drops precipitously, dramatically decreasing the inferred N$^{++}$/O$^{+}$. However, such densities would make the emergence of the resonant Lyman-$\alpha$ and Mg {\sc ii} difficult to explain, except in the complete absence of dust or if we are conveniently looking down an optically thin channel, and would imply super-solar O/H. Similarly, as discussed above, high densities would not explain the high N$^{++}$/O$^{++}$.

To remain conservative, for each $T_e$ value, we take the N$^{++}$/O$^{++}$ and N$^{++}$/O$^{+}$ values that yield the lowest nitrogen abundance, and combine these to obtain log N$^{++}$/(O$^{+}$ + O$^{++}$), treating this as our lower limit on the total N/O. Our fiducial case thus implies log (N/O$) > -0.25$, more than four times higher than solar (log (N/O$)_\odot = -0.86$; Table~\ref{tab:fiducial} and Figure~\ref{fig:fig1}). We obtain log (N/O$) > -0.13$ and log (N/O$) > -0.49$ in our conservative low and high temperature cases, while our ultra-high-temperature bounding case still yields log (N/O$) > -0.55$, twice that of the solar ratio.

The calculations presented here have not included the \NIV\ emission line. Considering this higher ionisation state only makes this picture more puzzling, since some \NIV\ emission can originate from the O$^{++}$ zone.
The weak He~{\sc ii} $\lambda$1486 emission and low derived N$^{3+}$/N$^{++}$ ratio (Table~\ref{tab:abund}) would seem to imply the O$^{3+}$ abundance is relatively low.
Thus, considering the N$^{3+}$ ion would likely only increase the inferred N/O ratio.

In summary, from measurements and limits on the \NIIIl1750~/~\OIIIsfll1660,1666, \NIIIl1750~/~\OIIIl4363, and \NIIIl1750~/~\OIIll3726, 3729 emission line ratios, we infer lower limits on the total N/O abundance ratio that most conservatively suggest log(N/O$) > -0.55$, which is twice the solar value, or, with more realistic assumptions, log(N/O) $>-0.25$, which is four times the solar value.
We will return to the surprising implications of this in Section~\ref{sec:discussion} and now repeat these arguments to derive carbon abundances.

\subsection{Constraints on C/O}
\label{sub:measure_co}

Given the detection of \CIIIll1909 reported by \citetalias{Bunker2023}, we can derive constraints on the C/O abundance ratio following the same set of assumptions and procedure outlined in Section~\ref{sub:measure_no} for N/O. We reiterate that, for each temperature, the reported abundance ratios are lower limits (see Section~\ref{sub:measure_no}).

As with nitrogen, ionisation potentials of carbon 
are such that the C$^{++}$ ionisation zone should overlap with the O$^{+}$ and O$^{++}$ zones. Therefore, we again assume that C$^{++}$/(O$^{+}$ + O$^{++}$) provides a lower limit on the total C/O ratio yielding $\log({\rm C/O})>-0.78$ in our fiducial case, and $\log({\rm C/O})>-0.95$ under more conservative assumptions (see Tables~\ref{tab:fiducial}~and~\ref{tab:abund}). These values are somewhat higher than previously reported in high-redshift galaxies \citep{ArellanoCordova2022, Jones2023}, but are reasonably consistent with lower-redshift objects (Figure~\ref{fig:fig1}).

\subsection{Constraints on O/H}
\label{sub:measure_oh}

To remain consistent with the approaches used in Sections~\ref{sub:measure_no} and \ref{sub:measure_co}, we derive $T_e$-based O/H values, adopting the range of temperatures assumed in Section~\ref{sub:temperature} for $T_e$(O{\sc iii}), and converting these into $T_e$(O{\sc ii}) using the calibration from \citet{Pilyugin2009}.
We derive O$^{++}$/H$^{+}$ from the ratio of \OIIIl4363 / H$\gamma$, and O$^{+}$/H$^{+}$ from the ratio of \OIIll3726, 3729 / H$\gamma$, and assume that the total oxygen abundance of GN-z11 is well approximated as O/H $\approx$ (O$^{++}$ + O$^{+}$)/H$^{+}$. Although the weak He {\sc ii} emission reported by \citetalias{Bunker2023} could suggest some fraction of oxygen is present as O$^{3+}$, the uncertainty in our measurement is more likely dominated by our inability to precisely constrain temperature.

Table~\ref{tab:abund} demonstrates the large temperature dependence of the total O/H ratio, which changes by almost two orders of magnitude across our adopted range.
Nonetheless, our fiducial temperature yields $12+\log(\rm O/H)=7.82$, broadly consistent with the value inferred by \citetalias{Bunker2023} from both strong-line and SED fitting methods. 

\section{Discussion} 
\label{sec:discussion}

The abundance ratios inferred from the bright nitrogen emission lines in the spectrum of GN-z11 imply a highly nitrogen-enriched ISM. In this section, we argue that such super-solar N/O ratios are particularly peculiar at $z=10.6$, and propose several scenarios that may explain this behaviour.

\subsection{Could GN-z11 be powered by a massive black hole?} \label{sec:discussion:AGN}

Although \NIIIl1750 has rarely been observed in star-forming galaxies, a `nitrogen-loud' population of quasars has been identified in SDSS exhibiting strong \NIIIl1750 and \NIVl1486 emission \citep{Jiang2008}. Furthermore, a recent spectrum of a $z=5.5$ AGN \citep{Ubler2023} also shows these nitrogen lines. \citetalias{Bunker2023} found that rest-frame UV emission line ratios in GN-z11 are generally more consistent with star-formation models \citep[e.g.][]{Feltre2016, Hirschmann2019}, but there is overlap with the parameter-space inhabited by some AGN models \citep{Nakajima2022_popiii}, and the possibility that GN-z11 hosts an AGN cannot not conclusively be ruled out (see also \citealt{Jiang2021}). It is unclear whether applying the $T_e$ method as outlined in Section~\ref{sec:abundances} to derive metal abundance ratios is valid in the case of an AGN, or how to interpret the emission line fluxes at hand if they arise from a high-density broad-line region, and we thus discuss here the possibility that GN-z11 is powered by a massive black hole.

Similar to GN-z11, nitrogen loud quasars have been suggested to arise due to enhanced nitrogen abundance and are rare, comprising only $\sim1$ \% of the SDSS quasar sample \citep{Jiang2008}. However, they are observed at much lower redshift with longer possible metal enrichment timescales, and whether the elevated N/O is simply tracing an increase in O/H via secondary enrichment \citep{Batra2014} or whether nitrogen is specifically enriched \citep{Araki2012, Kochanek2016, Matsuoka2017} remains debated.

The equivalent width (EW) ratio of EW(\NIII)/EW(\CIII) in GN-z11 would place it in the top $\sim2$\% of the \citet{Jiang2008} sample (already sampling only $\sim1$\% of SDSS quasars). Considering that AGN are expected to be rare at $z>10$ given the drop in the AGN luminosity function \citep{Kulkarni2019}, it would be interesting if GN-z11 is part of such a rare subpopulation of AGN and would imply that nitrogen-loud quasars dominate the AGN population in the early Universe. Even if GN-z11 is an AGN, explaining the nitrogen-loud behaviour would likely require super-solar N/O and N/C ratios. 

As we discuss below, explaining super-solar N/O and N/C is difficult with standard stellar nucleosynthesis models. One alternative to explain the rarity and the nitrogen-enrichment in such quasars is enrichment by tidal disruption events \citep[TDEs,][]{Kochanek2016}. As a star nears a supermassive black hole, it can be tidally disrupted. Since the core of intermediate-mass stars are rich in light elements, such TDEs can result in abundance anomalies with high N/C ratios (e.g. \citealt{Cenko2016, Yang2017, Brown2018, Sheng2021}) and help explain the emission patterns of nitrogen-rich quasars (\citealt{Kochanek2016}). 

To summarize, we cannot conclusively determine whether GN-z11 is a nitrogen-loud quasar and/or powered by a TDE, although its emission properties would put it amongst the rarest objects known in this category. Furthermore, the likelihood of such scenarios would have to be confronted quantitatively against the expected demographics of super-massive black holes which are expected to plummet at high redshift (e.g. \citealt{Volonteri2010}). Even in this case, GN-z11 may still require super-solar N/O ratios which are challenging to explain with typical galactic enrichment models (Section~\ref{sec:discussion:AGBs}). We note that deep high resolution NIRSpec spectroscopy of this object could help reveal the presence or absence of broadened lines and shed light on its nature.


\subsection{Is the over-abundance of light elements in GN-z11 from traditional evolved stars?} 
\label{sec:discussion:AGBs}

Under the traditional paradigm for galactic chemical evolution, oxygen (and other $\alpha$-elements) is enriched first via core-collapse supernova (CCSN), while carbon appears on a slightly longer timescale through both CCSN and winds from asymptotic giant branch (AGB) stars, and nitrogen lags behind mainly as a product of AGB stars (see e.g. \citealt{Nomoto2013, Kobayashi2023} for reviews). A massive progenitor for an AGB star ($\approx 6\, {\rm M}_{\odot}$) requires $\approx 50$ Myr to move off the main-sequence and enter the giant phase, where its winds will deposit (primarily) carbon and nitrogen in the surrounding gas. Given that the age of the Universe is only 440~Myr at $z=10.6$, this would put the birth of such AGB progenitors in a star formation burst at $z \geq 12$. 
Requiring a significant contribution from intermediate-mass progenitors ($\approx 3-4\, {\rm M}_{\odot}$) would push these requirements to even earlier times ($z\geq 14$). This is possible given that observed high-redshift Balmer Breaks may indicate star formation as early at 250~Myr after the Big Bang \citep{Hashimoto2018}; however, if the timescale gets moved too far back, we may enter a regime dominated by Population~III (Pop.~III) star formation (e.g. \citealt{Bromm2013}) where the IMF and yields may be different (see below).

While the timescales for AGB stars are reasonable, a crucial aspect of GN-z11 is the over-enrichment of nitrogen compared to oxygen. This implies efficient light element production, but also very inefficient oxygen enrichment by CCSNe given the fiducial metallicity of the object. This must apply both in the hypothetical first burst of star formation at $z\gtrsim 14$ giving rise to AGB progenitors, and in the current event at $z = 10.6$ powering the observed emission lines with star formation rate $\sim 20\ {\rm M}_{\odot} \, \text{yr}^{-1}$ (\citetalias{Bunker2023}, \citealt{Tachella2023}). A coincidental sequence of events could provide a mechanism to maintain the observed N/O and C/O. For example, the older star formation event could have either expelled most of the early oxygen in a powerful outflow or failed to produce it by collapsing most SNe progenitors directly into black holes. Next, AGBs enriched the gas in nitrogen over tens of Myrs, and we are catching GN-z11 just before the most recent CCSNe at $z=10.6$ enrich its ISM significantly in oxygen. 

Such a fine-tuned scenario to explain the observed N/O and C/O at $z=10.6$ through AGB enrichment is thus possible but rather contrived, and would need careful quantitative validation against models of galactic enrichment. In this case, most high-redshift luminous galaxies should be nitrogen-enriched. Stellar evolution models of AGB stars at low and very low metallicities exhibit shorter main-sequence lifetimes to reach the giant phase, and increased nitrogen and carbon production (e.g. \citealt{Cristallo2015, Ventura2021, Gil-Pons2022}). This could help quicken nitrogen and carbon enrichment timescales and relieve potential tensions, as would including more massive AGB and super-AGB progenitors that evolve quicker but whose yields remains uncertain (e.g. \citealt{Ventura2010, Siess2010, Doherty2014, Gil-Pons2022}; see \citealt{Karakas2014} for a review). Alternatively, stellar rotation and magnetic fields could also modify metal yields during the AGB phase (e.g. \citealt{Meynet2002, Piersanti2013, denHartogh2019}), but a consensus on the respective importance of these mechanisms for galactic-scale enrichment remains lacking.

The spectrum of GN-z11 also presents a tentative detection of the \HeII $\lambda$1486 line that could be associated with young, massive stars in the Wolf-Rayet phase. Such stars evolve quickly off the main-sequence ($2-3$ Myr at solar metallicity; \citealt{Meynet1995}) bypassing the longer timescales associated to AGB stars, and power winds that could contribute significant carbon and nitrogen to the chemical enrichment of the galaxy (see e.g. \citealt{Crowther2007, Vink2022} for reviews). Galaxies dominated by Wolf-Rayet features have been linked to elevated N/O ratios at lower redshifts (e.g. \citealt{Brinchmann2008, Berg2011, Masters2014}), although their reported N/O ratios ($\log({\rm N/O})\lesssim - 0.5$) remain much lower than reported here in GN-z11 ($\log({\rm N/O})\gtrsim - 0.3$). However, the weak \HeII line would imply limited contribution of Wolf-Rayet starlight to the integrated spectrum, and such helium-line ratios can also be explained by harder ionizing stellar populations at higher redshift (see e.g. discussion in \citealt{Steidel2016}), disfavouring this interpretation. It also remains unclear whether enough Wolf-Rayet stars could be present without CCSNe, which would rapidly enrich the ISM with oxygen and lower N/O and C/O. Nonetheless, if galaxies at $z\geq10$ commonly undergo a Wolf-Rayet-dominated phase, we would expect to see more systems with elevated N/O abundance ratios at high-redshift, that are yet to be detected but could be probed by further JWST observations (e.g. \citealt{RobertsBorsani2022, Cameron2023}).

To summarize, explaining the super-solar N/O and C/O abundance ratios in GN-z11 using traditional stellar evolutionary tracks is possible, but likely requires a very specific formation scenario. 


\subsection{Are we witnessing the chemical signatures of primordial or exotic stellar evolution channels?}
\label{sec:discussion:popiii}

Another possibility to explain the observed high N/O ratios in the ISM of GN-z11 is that the stars powering the bright emission lines at $z=10.6$ are rapidly enriching the ISM in nitrogen. Such a production mechanism would necessitate unusual stellar evolution channels, likely to be rare or cease to operate at later times (or both), as no metal-poor galaxies exhibit this level of nitrogen enhancement at lower redshifts (Figure~\ref{fig:fig1}). 

Furthermore, if this channel were to be common, its chemical signatures would likely be imprinted on the abundances of low-metallicity halo stars around our Milky Way. Carbon enhancements are commonly detected in halo stars (e.g. \citealt{Frebel2015} for a review). In contrast, nitrogen enhancements are rare (e.g. \citealt{Johnson2007, Pols2012, Simpson2019}), and are often attributed to binary mass-transfer at later times (e.g. \citealt{Suda2004, Pols2012, Fernandez-Trincado2019, Roriz2023}) rather than being set by the birth environment of the star. However, there are important examples where binary evolution is not the preferred explanation. In particular, HE~1327-2326 \citep{Frebel2008} and J0815+4729 \citep{Gonzalez2020} exhibit no signatures of mass transfer and binary companions, despite their drastic enhancements of carbon and nitrogen \citep{Aoki2006}. Many have thus proposed that the abundances of HE~1327-2326 were set at high redshift, possibly by Population~III (Pop.~III) primordial stars \citep{Iwamoto2005,Frebel2005,Hirschi2007,Heger2010,Ezzeddine2019}.

Given the parallels with such nitrogen-enhanced metal-poor stars, we now assess the likelihood that in-situ Pop.~III star formation could be responsible for the observed abundance ratios of GN-z11. To this end, we explore a compilation of Pop.~III and low-metallicity SN yields, scanning across the available parameter space to identify models that produce abundance ratios close to our derived fiducial values. Namely, we look for (i) a carbon to oxygen ratio such that $\log({\rm C/O})>-0.78$, (ii) a high ratio of nitrogen to oxygen ($\log({\rm N/O})>-0.25$), (iii) more nitrogen than carbon ($\log({\rm N/C})>0.53$), and (iv) a significant amount of nitrogen mass per event (at least $0.01\ {\rm M_{\odot}}$). These thresholds correspond to our fiducial model (see Table~\ref{tab:abund}).

We consider Pop.~III SN yields from \cite{Heger2010} that include metal production for various stellar mass progenitors, SNe energies, piston location, and mixing amounts. We include yields from \cite{Takahashi2018} that further account for stellar rotation and magnetic fields, in addition to progenitor mass. Finally, we search the yields of more metal-enriched rotating stars from \cite{Limongi2018} where the rotation is known to enhance the nitrogen abundance. While this compilation of yields is by no definition complete, they span the range of relevant physical mechanisms that could help explain the observed abundances of GN-z11.

Within the \cite{Heger2010} data set we find stars with a mass of $25\ {\rm M_{\odot}}$ or $39\ {\rm M_{\odot}}$, with explosion energies of $0.3\times10^{51}\ {\rm erg}$ and $0.6\times10^{51}\ {\rm erg}$, respectively, exhibit the required abundance patterns. No models amongst those presented in \cite{Takahashi2018} that satisfy our criteria. For the yields from \cite{Limongi2018}, we find two stars\footnote{If we switch our constraints to the most lenient region of our allowed parameter space, only one additional star, a $40\ {\rm M_{\odot}}$ star at $0.1\ Z_{\odot}$ rotating at 150~km/s fits our criteria. Likewise, adopting more conservative thresholds does not change our results for the \cite{Heger2010} or \cite{Takahashi2018} data sets.}: an $80\ {\rm M_{\odot}}$ non-rotating star at $0.1\ Z_{\odot}$ and an $80\ {\rm M_{\odot}}$ star rotating at 300~km/s at $0.001\ Z_{\odot}$. We emphasize here that by no means is this search a reflection of the accuracy of these stellar evolution calculations, rather it is an exercise to determine whether Pop.~III SN, rotating stars, or faint SN have the potential to explain GN-z11 or whether other physics is required. 

Our search demonstrates that in certain cases, the abundance patterns observed in GN-z11 can be reproduced both by certain Pop.~III SN as well as more metal-enriched models. However, we note that within the parameter space of possible SN explosions, only a select few models are able to reproduce the abundances which results in a fine-tuning problem. It is highly unlikely that every star present in the metal enriched environment of GN-z11 is exactly $80\ {\rm M_{\odot}}$. Furthermore, why should the Pop.~III stars that potentially enriched GN-z11 only appear at $25\ {\rm M_{\odot}}$ or $39\ {\rm M_{\odot}}$? GN-z11 is bright and a significant amount of nitrogen is required to produce the observed luminosity. Each of these explosions produces $<0.1\ {\rm M_{\odot}}$ of nitrogen, so a significant number are needed in order to enrich the galaxy to levels where emission lines are detectable. It is unclear how so many stars could form at a particular mass or how to maintain Pop.~III star formation for such an extended period of time. Similarly, if Pop.~III stars really formed at particular masses, nitrogen enrichment in the stellar halo might not be so uncommon. 
We disfavour solutions of similar ilk, e.g. Pop.~III stellar winds that can similarly produce the correct abundances but require fine-tuning \citep{Hirschi2007}.

In summary, similar to the AGB wind scenario presented above, faint Pop.~III SN or low-metallicity rotating stars have the capability of producing the yields reported for GN-z11; however, the scenario is fine-tuned and significant deviations from the local stellar IMF would be required for such abundance ratios as observed in GN-z11 to manifest.


\subsection{Are we observing stellar encounters in dense star clusters?}
\label{sec:discussion:stellarcluster}

Following the apparent fine-tuning required to explain the abundance ratios of GN-z11 through specific stellar evolution mechanisms, we explore an alternative scenario where the metal content results from dynamical processes within the particular environment of GN-z11. More specifically, runaway stellar collisions in dense early star clusters could provide a high-redshift-only, rare mechanism to elevate nitrogen production that fits the compact morphology and high star-formation rate observed in GN-z11 (\citealt{Tachella2023}).

Within dense stellar clusters, high-mass stars can rapidly sink to the centre due to mass segregation (e.g. \citealt{PortegiesZwart2002, Gurkan2004}) and collide. If the cluster is dense enough, multiple collisions can occur on time scales shorter than the main-sequence lifetimes of massive stars, thus before the first SNe explode, and form very massive stars in its centre (\citealt{PortegiesZwart1999}). This scenario is much more likely to occur in high-redshift galaxies due to the higher gas densities and increased merger rates and can provide a mechanism for the production of massive black hole seeds to explain the origins of high-redshift supermassive black holes (e.g. \citealt{Katz2015}).

In the case of GN-z11, the presence of such very massive stars embedded in a star cluster could help explain its abundance ratios. Massive stars produced by runaway collisions are may be well-mixed (e.g. \citealt{Gaburov2008}), bringing light elements towards their surface. While at low metallicities the very massive star is either expected to collapse to a black hole with minimal mass loss or explode as a pair-instability SN (the outcome depends on mass), metal enriched very massive stars are expected to host powerful stellar winds (\citealt{Vink2022}). This can lead to fast and efficient enrichment of light elements such as carbon and nitrogen at the expense of oxygen (\citealt{Glebbeek2009}). Furthermore, the number of SNe is reduced in such a formation scenario as most SN progenitors merge quickly into a single object, reducing the production of carbon and oxygen and helping to increase N/O.

It remains unclear whether a single cluster undergoing collisional runaway produces enough metal and nitrogen mass to power the emission lines of GN-z11. However, multiple massive stellar clusters are expected to form simultaneously if the galaxy is undergoing an external trigger inducing strong compressive tides (e.g. a merger; \citealt{Renaud2015, Li2017}), a process best observed in the Antennae galaxies (\citealt{Bastian2009}). Furthermore, not all stellar clusters need to undergo runaway collisions to create the observed spectrum of GN-z11 -- star clusters where collisional runaway is efficient could be responsible for the nitrogen emission lines, whereas the oxygen and carbon emission lines can be spread throughout the other, more classical star clusters of the galaxy. Subtly different conditions in each cluster could thus participate in driving enhanced N/O integrated over the galaxy.

While the winds of a collisionally-produced very massive star may produce the nitrogen observed in GN-z11, one of the limitations of this scenario is the fact that the very massive star must form quickly, before the stars can explode via SN. However, regardless of whether this process ensues, some massive stars in the cluster may collapse into stellar mass black holes (i.e. $<100\ {\rm M_{\odot}}$). The dense environment of the stellar cluster could favour close encounters between stars and black holes resulting in TDEs, which could help explain the nitrogen and helium emission (e.g. \citealt{Kochanek2016} and discussion in Section~\ref{sec:discussion:AGN}). This parallels the TDE scenario for AGN but uses lower mass black holes.

To summarize, this collisional runaway scenario evokes exotic dynamical processes much less likely to occur at the lower densities of the lower-redshift Universe, and thus fits the rarity and peculiarity of GN-z11 without modifying our base understanding of stellar evolution. There are however large remaining uncertainties associated with modelling stellar evolution during runaway collisions (e.g. mixing during stellar collisions, the nucleosynthesis and stellar winds associated to the central massive star), as well as how much nitrogen, carbon, and oxygen are released during a TDE. Nonetheless, these findings strongly advocate future theoretical studies exploring and testing these scenarios quantitatively.

\section{Summary} \label{sec:conclusion}

Based on its luminosity alone, GN-z11 is a remarkable object at $z>10$ (\citealt{Oesch2015, Oesch2016}). Its recent spectroscopic follow-up in \citetalias{Bunker2023} further revealed the extent of this peculiarity, highlighting emission lines from carbon, oxygen and nitrogen amongst others, and showcasing the power of \emph{JWST}/NIRSpec spectroscopy to characterize the physical properties of galaxies less than $500$ Myr after the Big Bang. 

In particular, the presence of strong \NIIIl 1750 and \CIIIll1909 emission lines allows for unprecedented constraints on chemical abundance ratios, and the high \NIIIl 1750 / \OIIIsfll 1660, 1666 ratio could imply unusually high N/O (\citetalias{Bunker2023}). In this paper, we quantitatively derive the abundance ratios implied by these emission line fluxes and find $\log({\rm N/O})>-0.25$, $\log({\rm C/O})>-0.78$, and $\log({\rm O/H})\approx7.82$ for our fiducial model. This indicates super-solar nitrogen enrichment in GN-z11 within the first $\sim$440 Myr of cosmic history. More conservative assumptions in our modelling suggest $\log({\rm N/O})>-0.49$, $\log({\rm C/O})>-0.95$ and $\log({\rm O/H})\lesssim8.60$, still yielding a super-solar N/O. 

We explore how our derived values vary with different assumptions of temperature, density, dust, and ionisation corrections, finding that none of these can reasonably explain the high \NIIIl 1750 / \OIIIsfll 1660, 1666 ratio without invoking a high N/O ratio. Given the longer enrichment timescales typically associated with nitrogen  compared to oxygen, this over-enrichment is highly unexpected and seemingly at odds with the young age of the Universe at $z=10.6$.

We review whether the emission pattern observed in GNz11 could be powered by an AGN, disfavoured in \citetalias{Bunker2023}, but which could bias the inferred N/O. We find qualitative parallels between this object and the population of rare `nitrogen-loud' quasars, although emission line ratios observed in GNz11 would put it as an outlier of this already-rare population. We cannot conclusively exclude this scenario, but note that the mechanisms invoked to explain these nitrogen-loud objects either involve significant nitrogen enrichment or tidal disruption events, both of which have interesting implications at $z=10.6$. 

Assuming instead that GN-z11 is indeed a star-forming galaxy, as preferred by \citetalias{Bunker2023}, we then review stellar processes that could produce high N/O at such early cosmic times. Traditional models of nitrogen-enrichment from AGB winds would likely require a highly contrived formation scenario, which cannot be ruled out but requires extensive validation against quantitative galactic enrichment models. Similarly, metal yields from exotic stellar evolution channels, including rotating and Pop.~III massive stars, generally disfavour high nitrogen-to-oxygen production. Individual progenitor models can lead to high N/O, but generalizing across the galaxy would require an extremely finely-tuned progenitor mass function and initial conditions. 

Lastly, we explore whether exotic dynamical mechanisms operating at high redshift could explain the apparent nitrogen-enhancement in GN-z11. Runaway stellar collisions in the cores of dense, high-redshift stellar clusters can lead to the formation of very massive stars, leading to rapid and abundant nitrogen production and an under-production of oxygen. There are large quantitative uncertainties with this scenario, but it provides an avenue to simultaneously explain the high N/O in GN-z11 and the lack of low-redshift counterparts where gas densities become lower. These same star clusters would also be ideal sites to host TDEs which could also lead to nitrogen enhancements as discussed for the AGN scenario.

Ultimately, we cannot conclusively distinguish between these scenarios and acknowledge our proposed list is unlikely to be exhaustive. Rather, the prominent and unusual \NIIIl1750 emission observed in GN-z11 should stimulate further studies that both quantitatively establish the likelihood of our proposed options and explore additional models that could explain such high N/O at $z=10.6$. 

Nonetheless, the fact that one of the first emission spectra observed at $z>10$ reveals such prominent nitrogen emission, uncommon at lower redshifts, suggests that we are only just opening a new frontier. \emph{JWST}/NIRSpec spectroscopic programs targeting larger samples of high-redshift galaxies will allow us to quantify the frequency of such bright \NIIIl1750 emission amongst the $z>10$ population and refine our understanding of how the first metals appeared in the Universe.

\section*{Acknowledgements}

We thank James Matthews for helpful discussions in relation to this work.
AJC and AS have received funding from the European Research Council (ERC) under the European Union’s Horizon 2020 Advanced Grant 789056 ``First Galaxies''. MR and HK are supported by the Beecroft Fellowship funded by Adrian Beecroft.
CHIANTI is a collaborative project involving George Mason University, the University of Michigan (USA), University of Cambridge (UK) and NASA Goddard Space Flight Center (USA). 
For the purpose of Open Access, the author has applied a CC BY public copyright licence to any Author Accepted Manuscript version arising from this submission.

\section*{Data Availability}

Emission line fluxes measured in GN-z11 are publicly available in \citet{Bunker2023}. Derived abundance ratios from these fluxes are available in Table~\ref{tab:abund}.



\bibliographystyle{mnras}
\bibliography{paper_bib} 




\appendix

\section{Full table of abundance calculation results.}

In Table~\ref{tab:abund} we present the full range of ion abundance ratios we obtain from the calculations presented in Section~\ref{sec:abundances}.

{\renewcommand{\arraystretch}{1.6}
\begin{table*}
    \centering
    \begin{tabular}{clcccclc}
\hline
\hline
Row & Abundance ratio & $T_e = 1.05 \times 10^4$ K & $T_e = 1.46 \times 10^4$ K  & $T_e = 2.36 \times 10^4$ K  & $T_e = 3.0 \times 10^4$ K &  Notes \\
 &  & $n_e=100$ cm$^{-3}$ & $n_e=100$ cm$^{-3}$ & $n_e=100$ cm$^{-3}$ & $n_e=100$ cm$^{-3}$ & \\
 &  &  & \textbf{Fiducial case} &  & & \\
\hline
\hline
1 & log $\frac{{\rm N}^{++}}{{\rm O}^{++}}$ & $-$0.12 & \textbf{$-$0.07} & $-$0.02 & 0.00 & From \NIII\  / \OIIIsfll 1660, 1666$^\dag$ \\
2 & log $\frac{{\rm N}^{++}}{{\rm O}^{++}}$ & 0.04 & \textbf{$-$0.19} & $-$0.42 & $-$0.48 & From \NIII\  / \OIIIl 4363$^\ddag$ \\
3 & log $\frac{{\rm N}^{++}}{{\rm O}^{+}}$ & 1.69 & \textbf{1.08} & 0.55 & 0.38 & $T_e$(O{\sc ii}) from \citet{Pilyugin2009}$^\ddag$ \\
4 & log $\frac{{\rm N}^{++}}{{\rm O}^{+}}$ & 0.96 & \textbf{0.64} & 0.33 & 0.23 & $T_e$(O{\sc ii}) $=0.7\times T_e$(O{\sc iii})$^\ddag$ \\
\hline
5 & log $\frac{{\rm N}^{++}}{({\rm O}^{+}+{\rm O}^{++})}$ & $-$0.13 & \textbf{$-$0.25} & $-$0.49 & $-$0.55 & $^*$ \\
\hline
\hline
6 & log $\frac{{\rm C}^{++}}{{\rm O}^{++}}$ & $-$0.73 & \textbf{$-$0.6} & $-$0.48 & $-$0.44 & From \CIII\  / \OIIIsfll 1660, 1666$^\dag$ \\
7 & log $\frac{{\rm C}^{++}}{{\rm O}^{++}}$  & $-$0.57 & \textbf{$-$0.72} & $-$0.88 & $-$0.92 & From \CIII\  / \OIIIl 4363$^\ddag$ \\
8 & log $\frac{{\rm C}^{++}}{{\rm O}^{+}}$ & 1.08 & \textbf{0.55} & 0.08 & $-$0.06 & $T_e$(O{\sc ii}) from \citet{Pilyugin2009}$^\ddag$ \\
9 & log $\frac{{\rm C}^{++}}{{\rm O}^{+}}$ & 0.35 & \textbf{0.11} & $-$0.13 & $-$0.21 & $T_e$(O{\sc ii}) $=0.7\times T_e$(O{\sc iii})$^\ddag$ \\
\hline
10 & log $\frac{{\rm C}^{++}}{({\rm O}^{+}+{\rm O}^{++})}$  & $-$0.74 & \textbf{$-$0.78} & $-$0.95 & $-$0.99 & $^*$ \\
\hline
\hline
11 & log $\frac{{\rm N}^{++}}{{\rm C}^{++}}$ & 0.61 & \textbf{0.53} & 0.46 & 0.44 &  \\
12 & log $\frac{{\rm N}^{3+}}{{\rm N}^{++}}$ & $-$0.07 & \textbf{$-$0.17} & $-$0.25 & $-$0.26 & $T_e$(N{\sc iv}) $= T_e$(N{\sc iii}) \\
13 & log $\frac{{\rm O}^{++}}{{\rm O}^{+}}$ & 1.68 & \textbf{1.28} & 0.94 & 0.82 & $T_e$(O{\sc ii}) from \citet{Pilyugin2009} \\
\hline
14 & log $\frac{{\rm O}^{++}}{{\rm H}^{+}}$ & $-$3.41 & \textbf{$-$4.2} & $-$5.02 & $-$5.33 & \\
15 & log $\frac{{\rm O}^{+}}{{\rm H}^{+}}$ & $-$5.09 & \textbf{$-$5.48} & $-$5.96 & $-$6.15 & $T_e$(O{\sc ii}) from \citet{Pilyugin2009} \\
16 & log $\frac{{\rm O}}{{\rm H}}$ & $-$3.4 & \textbf{$-$4.18} & $-$4.97 & $-$5.27 & \\
17 & 12 + log $\frac{{\rm O}}{{\rm H}}$ & 8.6 & \textbf{7.82} & 7.03 & 6.73 & \\
\hline
18 & log $\frac{{\rm N}^{++}}{{\rm H}^{+}}$ & $-$3.43 & \textbf{$-$4.45} & $-$5.49 & $-$5.86 & \\
19 & log $\frac{{\rm N}^{3+}}{{\rm H}^{+}}$ & $-$3.51 & \textbf{$-$4.62} & $-$5.74 & $-$6.12 & \\
20 & log $\frac{({\rm N}^{++}+{\rm N}^{3+})}{{\rm H}^{+}}$ & $-$3.17 & \textbf{$-$4.22} & $-$5.3 & $-$5.67 & \\
21 & log $\frac{{\rm C}^{++}}{{\rm H}^{+}}$ & $-$4.04 & \textbf{$-$4.98} & $-$5.95 & $-$6.31 & \\
\hline
\hline
    \end{tabular}
    \caption{Abundance ratios calculated for $n_e=100$ cm$^{-3}$ under a range of temperature assumptions. The temperature given in the header row is the adopted $T_e$(N{\sc iii}), $T_e$(C{\sc iii}) and $T_e$(O{\sc iii}) for that column. $T_e$(O{\sc ii}) inferred in two different ways, either adopting the calibration from \citet{Pilyugin2009}, which yields $T_e$(O{\sc ii}) $=$ [1.14, 1.48, 2.24, 2.77] $\times 10^4$ K for each column, or taking the more exaggerated assumption that $T_e$(O{\sc ii}) $=0.7\times T_e$(O{\sc iii}) which gives $T_e$(O{\sc ii}) $=$ [0.74, 1.02, 1.65, 2.10] $\times 10^4$ K. \\
    $^\dag$ Values in this row are a lower limit on the abundance ratio, since the \OIIIsfll 1660, 1666 value used is the $2\sigma$ upper limit reported in \citetalias{Bunker2023}. \\
    $^\ddag$ Values in this row can be thought of as a lower limit on the abundance ratio, since it was computed assuming no dust reddening, and invoking the presence of dust would preferentially boost the shorter wavelength line in our calculation. \\
    $^*$ This row gives the minimum possible X$^{++}$/(O$^{+}$ + O$^{++}$) ratio that can be obtained from summing the possible X$^{++}$/O$^{+}$ and X$^{++}$/O$^{++}$ values from each column (where X is nitrogen or carbon, as per the `abundance ratio' column). \\ 
    For reference, solar values are: $\log ({\rm N/O})_\odot = -0.86$, $\log ({\rm C/O})_\odot = -0.26$, $12+\log ({\rm O/H})_\odot = 8.69$ \citep{Asplund2009}. \\
    }
    \label{tab:abund}
\end{table*}
}


\bsp	
\label{lastpage}
\end{document}